\documentclass[a4paper,10pt]{article}
\pdfoutput=1 

\usepackage{jcappub} 

\usepackage{hyperref}

\usepackage[T1]{fontenc} 
\usepackage{footnote}
\def\be{\begin{equation}}
\def\ee{\end{equation}}
\def\ba{\begin{eqnarray}}
\def\ea{\end{eqnarray}}
\newcommand{\dd}{\mathrm{d}}

\usepackage{multirow}

\usepackage{array}

\title{\boldmath{Testing extended Jordan-Brans-Dicke theories with future cosmological observations}}

\author[a,b]{M. Ballardini}
\author[c]{D. Sapone}
\author[d,e,f]{C. Umilt\`a}
\author[b,g]{F. Finelli}
\author[b,g]{and D. Paoletti}

\affiliation[a]{Department of Physics and Astronomy, University of the Western Cape, Cape Town 7535, South Africa}
\affiliation[b]{INAF/OAS Bologna, via Gobetti 101, I-40129 Bologna, Italy}
\affiliation[c]{Cosmology and Theoretical Astrophysics group, Departamento de F\'isica, 
FCFM, Universidad de Chile, Blanco Encalada 2008, Santiago, Chile}
\affiliation[d]{University of Cincinnati, Cincinnati, Ohio 45221}
\affiliation[e]{Institut d'Astrophysique de Paris, CNRS (UMR7095), 98 bis Boulevard 
Arago, F-75014, Paris, France}
\affiliation[f]{Sorbonne Universit\'es, Institut Lagrange de Paris (ILP), 98 bis Boulevard 
Arago, 75014 Paris, France}
\affiliation[g]{INFN, Sezione di Bologna, Via Berti Pichat 6/2, I-40127 Bologna, Italy}

\emailAdd{mario.ballardini@gmail.com}
\emailAdd{domenico.sapone@uchile.cl}
\emailAdd{umilta@ucmail.uc.edu}
\emailAdd{fabio.finelli@inaf.it}
\emailAdd{daniela.paoletti@inaf.it}

\abstract{The extended Jordan-Brans-Dicke (eJBD) theory of gravity is constrained by
a host of astrophysical and cosmological observations spanning a wide range of scales.
The current cosmological constraints on the first post-Newtonian parameter in these
simplest eJBD models in which the recent acceleration of the Universe is connected 
with the variation of the effective gravitational strength are consistent, but 
approximately two order of magnitude larger than the time-delay test within the Solar 
System. We forecast the capabilities of future galaxy surveys in combination with 
current and future CMB anisotropies measurements to further constrain the simplest 
dark energy models within eJBD theory of gravity. By considering two cases of a 
monomial potential (a quartic potential or a cosmological constant), we show how 
Euclid-like galaxy clustering and weak lensing data in combination with BOSS and 
future CMB observations have the potential to reach constraints on the first 
post-Newtonian parameter $\gamma_{\rm PN}$ comparable to those from the Solar System.}

\begin{document}
\maketitle
\flushbottom

\section{Introduction}
\label{sec:intro}
Jordan-Brans-Dicke (JBD) theory of gravity \cite{Jordan:1949zz,Brans:1961sx} is among the 
simplest extensions of general relativity (GR), in which the gravitational field is mediated by 
a scalar field whose inverse plays the role of an effective gravitational constant which varies 
in space and time.
JBD theory depends on just one additional parameter $\omega_\mathrm{BD}$, connected to the 
post-Newonian parameter $\gamma_\mathrm{PN} = (1+\omega_\mathrm{BD})/(2+\omega_\mathrm{BD})$ 
measuring the deviations from Einstein GR, which is recovered in the limit of 
$\gamma_\mathrm{PN} \rightarrow 1$, i.e. $\omega_\mathrm{BD} \rightarrow + \infty$.
Observations on a wide range of scales constrain JBD theory around GR: the tightest limits, 
$\gamma_\mathrm{PN} - 1 = (2.1 \pm 2.3) \times 10^{-5}$ (68\% CL) are obtained 
from radar timing data by the Cassini spacecraft within our Solar System \cite{Bertotti:2003rm}.

Extended JBD (eJBD) theory of gravity with a potential term for the scalar field:
\be
{\cal S} = \int d^4x \sqrt{-g}\, \left[\frac{1}{16 \pi}
\left( \phi R  - \frac{\omega_\mathrm{BD}}{\phi} g^{\mu \nu} \partial_{\mu} \phi \partial_{\nu} \phi \right) - V(\phi)
+ \mathcal{L}_{\rm m} \right] \,,
\label{eJBD}
\ee
include the simplest scalar-tensor models of dark energy in which the current acceleration 
of the Universe is connected to a variation of the effective gravitational constant 
\cite{Wetterich:1987fk,Uzan:1999ch,Perrotta:1999am,Bartolo:1999sq,Amendola:1999qq,Chiba:1999wt,Boisseau:2000pr} 
(see also Ref.~\cite{Cooper:1982du}).
These models are also known as extended quintessence \cite{Perrotta:1999am,Chiba:1999wt}.

The phenomenology in the eJBD theory of gravity is much richer than in Einstein Gravity (EG), since cosmological 
variation of the effective gravitational constant could lead to different predictions not only 
for cosmic microwave background (CMB) anisotropy \cite{Chen:1999qh} and the growth of structures, but also for 
Big Bang Nucleosynthesis (BBN) \cite{Copi:2003xd,Bambi:2005fi}. 

Testing the viability of the cosmology in eJBD theory is fully complementary to the Solar System 
constraints just presented.
For models described by Eq.~\eqref{eJBD} with a quadratic potential \cite{Cooper:1982du,Wetterich:1987fk,Finelli:2007wb}, 
the recent $Planck$ 2015 \cite{Aghanim:2015xee,Ade:2015zua} and baryonic acoustic oscillations 
(BAO) data \cite{Beutler:2011hx,Ross:2014qpa,Anderson:2013zyy} constrain 
$1 - \gamma_\mathrm{PN} < 0.003$ (95\% CL) \cite{Ballardini:2016cvy} 
(see also \cite{Ooba:2016slp} for constraints obtained by relaxing the hypothesis of flat spatial 
sections and \cite{Avilez:2013dxa,Li:2013nwa,Umilta:2015cta} for the constraints based on the 
$Planck$ 2013 data).
These cosmological constraints on $\gamma_{\rm PN}$ are approximately two order of magnitude looser 
than Solar System constraints. 

In this paper we investigate the capabilities of future CMB and large scale structures (LSS) 
observations to further improve the cosmological constraints on the post-Newtonian parameter 
$\gamma_{\rm PN}$ within the eJBD theory, as also forecasted in \cite{Acquaviva:2007mm,Alonso:2016suf}.
We expect that upcoming galaxy surveys such as 
DESI \footnote{\href{http://desi.lbl.gov/}{http://desi.lbl.gov/}} \cite{Levi:2013gra}, 
Euclid \footnote{\href{http://sci.esa.int/euclid/}{http://sci.esa.int/euclid/}} 
\cite{Laureijs:2011gra,Amendola:2012ys}, 
LSST \footnote{\href{http://www.lsst.org/}{http://www.lsst.org/}} \cite{Abell:2009aa}, 
SKA \footnote{\href{http://www.skatelescope.org/}{http://www.skatelescope.org/}}
\cite{Maartens:2015mra,Bacon:2018dui}, will help in improving the
constraints of structure formation on $\gamma_{\rm PN}$ for the eJBD theory. 
As a representative example of what we will gain from upcoming galaxy surveys, we consider the 
two main probes of Euclid, galaxy clustering (GC), and weak lensing (WL).
In addition, we will consider the role of possible future CMB polarization anisotropy observations, 
as AdvACT \cite{Calabrese:2014gwa}, CORE \cite{Delabrouille:2017rct,DiValentino:2016foa,Finelli:2016cyd}, 
LiteBIRD \cite{Matsumura:2013aja,Errard:2015cxa}, and S4 \cite{Abazajian:2016yjj}, in further improving 
on the $Planck$ measurements.

Our paper is organized as follows. After this introduction, we give a lighting review of 
eJBD recast as Induced Gravity (IG) (by a redefinition of the scalar field with standard units and 
standard kinetic term) in Section~\ref{sec:two}.
In Section~\ref{sec:three} and \ref{sec:four} we present the Fisher methodology for CMB and LSS 
for our science forecasts.
In Section~\ref{sec:five} we present our results and in Section~\ref{sec:conclusion} we draw our conclusions.

\section{Dark Energy within the extended Jordan-Brans-Dicke theories}
\label{sec:two}
\begin{figure}
\centering
\includegraphics[width=\textwidth]{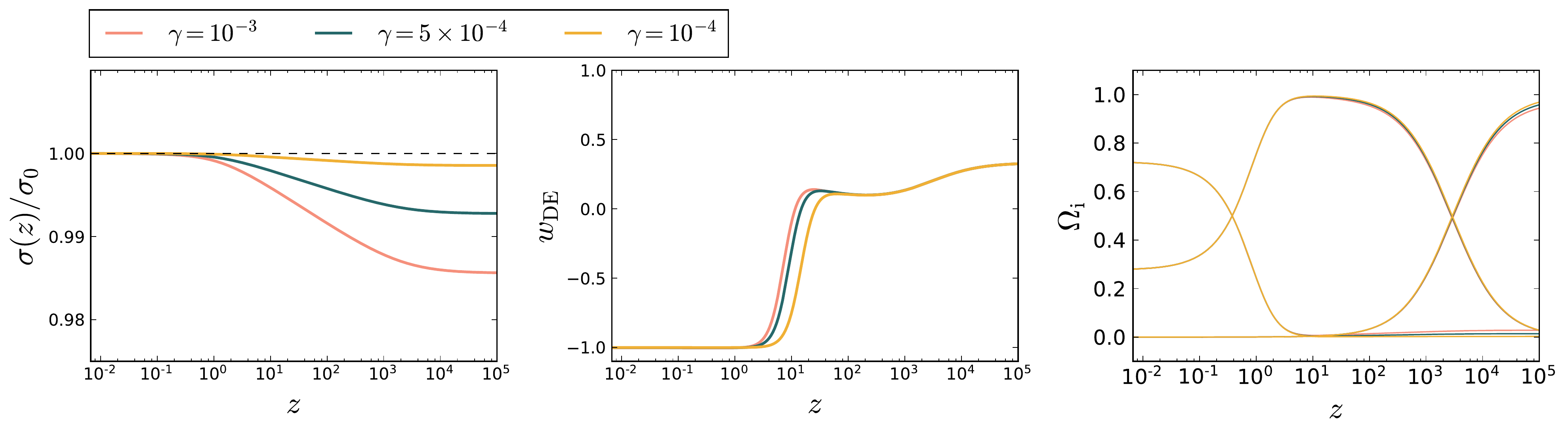}
\caption{Evolution of $\sigma/\sigma_0$ (top panel), $w_{\rm DE}$ (middle panel), 
and $\Omega_i$ (bottom panel) as function of $z$ for different choices of $\gamma$ 
from $\gamma=10^{-4}$ to $\gamma=10^{-3}$ for $n=4$. 
The value $\sigma_0$ of the scalar field at present is fixed consistently with the 
Cavendish-type measurement of the gravitational constant $G = 6.67\times10^{-8}$ cm$^3$ g$^{-1}$ s$^{-2}$.}
\label{fig:back}
\end{figure}
In this section we review some general considerations of the late-time cosmology within the 
eJBD theories. 

We consider a field redefinition to recast the eJBD action in Eq.~\eqref{eJBD} into an 
action for induced gravity (IG) with a standard kinetic term for a scalar field $\sigma$:
\be
\label{eqn:action}
\mathcal{S} = \int d^4x \sqrt{-g}\left[ \frac{\gamma\sigma^2R}{2}
- \frac{g^{\mu\nu}}{2}\partial_\mu\sigma\partial_\nu\sigma - V(\sigma) + \mathcal{L}_\mathrm{m} \right] \,.
\ee
where $\gamma = (4 \omega_\mathrm{BD})^{-1}$ and $\gamma\sigma^2=\phi$.

The cosmology evolution after inflation can be divided roughly in three stages and is summarized 
in Fig.~\ref{fig:back}. 
In the first stage relevant for our study, i.e. deep in the radiation era, $\sigma$ is almost frozen, 
since it is effectively massless and non-relativistic matter is subdominant.
During the subsequent matter dominated era, $\sigma$ is driven by non-relativistic matter to higher 
values, leading to an effective gravitational constant $G_{\rm N}(a)=1/\left(8\pi\gamma\sigma^2\right)$ 
which decrease in time. 
The potential $V(\sigma)$ kicks in only at recent times determining the rate of the accelerated 
expansion. 
For a simple monomial potential $V(\sigma) \propto \sigma^{n_{\rm IG}}$ and 
in absence of matter, exact power-law solutions for the scale factor $a (t) \sim t^p$ describing 
an accelerated expansion exist for the class of monomial potentials with 
$p = 2\frac{1+ (n_{\rm IG}+2) \gamma}{(n_{\rm IG}-4)(n_{\rm IG}-2) \gamma}$ 
\cite{Barrow:1990nv,Cerioni:2009kn}. A de Sitter solution for 
the scale factor is found instead for $n_{\rm IG}=2 \,, 4$.

In Fig.~\ref{fig:back} we display different quantities as a function of redshift:
the scalar field normalized to its value at present (left panel), the parameter of state $w_\mathrm{DE}$ 
of the effective dark energy component (middle panel), and the critical densities corresponding
to EG with a gravitational constant given by the current value of the scalar field, i.e.
$8 \pi G_{\rm N} (z=0)=1/(\gamma \sigma_0^2)$.
It is interesting to note from $w_\mathrm{DE}$ displayed in Fig.~\ref{fig:back} that the effective parameter 
of state for dark energy in these extended JBD models is similar to the so called old 
\cite{Doran:2006kp} and new \cite{Poulin:2018cxd} early dark energy models.

Since now on we will restrict ourselves to two cases of monomial potentials, 
i.e. $V(\sigma) \propto \sigma^{n_{\rm IG}}$ with $n_{\rm IG}=4$ or $n_{\rm IG}=0$,
suitable to reproduce a background cosmology in agreement with observations. We consider a scalar
field $\sigma=\sigma_i$ nearly at rest deep in the radiation era, since an initial non-vanishing time derivative
would be otherwise rapidly dissipated \cite{Finelli:2007wb}. The initial time derivative of the scalar field 
is taken as $d \sigma/\dd \tau = 3 \gamma \omega \sigma_i/2$ - 
with $\omega = \frac{\rho_{m \, 0}}{\sqrt{3 \gamma \rho_{r \, 0}} (1+6 \gamma) \sigma_i}$ - satisfying the 
equation of motion. We choose $\sigma_i$ by fixing the value $\sigma_0$ of the scalar field at present
consistently with the Cavendish-type measurement of the gravitational constant $G = 6.67\times10^{-8}$ cm$^3$ g$^{-1}$ s$^{-2}$, 
i.e. $\gamma \sigma_0^2 = \frac{1}{8\pi G}\frac{1 + 8\gamma}{1 + 6\gamma}$.
We also consider adiabatic initial conditions for fluctuations \cite{Paoletti:2018xet}.
In this way for a given potential the models we study have just one parameter in addition to the $\Lambda$CDM model, 
i.e. the coupling to the Ricci curvature $\Lambda$CDM model $\gamma$. 

The evolution of linear perturbations in this class of eJBD can be described with a set of dimensionless 
functions $\alpha_{\rm M} = \dd \ln \phi / \dd \ln a$, $\alpha_{\rm B} = -\alpha_{\rm M}$, 
$\alpha_{\rm K} = \omega_{\rm BD} \alpha_{\rm M}^2$, and $\alpha_{\rm T} = 0$ 
according to the parametrisation introduced in Ref.~\cite{Bellini:2014fua}.

$Planck$ 2015 temperature, polarization and lensing \cite{Aghanim:2015xee,Ade:2015zua} 
constrain $\gamma < 0.0017$ at 95\% CL 
and by combining with BAO data the 95\% CL upper bound tightens to $0.00075$ 
\cite{Ballardini:2016cvy}. 
The cosmological variation of the effective gravitational strength between now respect to the 
one in the relativistic era is constrained as $|\delta G_N/G_N| < 0.039$ at 95\% CL 
\cite{Ballardini:2016cvy}.
Such eJBD models predict a value for the Hubble parameter larger than $\Lambda$CDM, because of 
a degeneracy between $\gamma$ and $H_0$. This effect can be easily understood by interpreting the 
larger value of the effective gravitational constant in the past as a larger 
number of relativistic degrees of freedom.

Constraints on $\gamma$ and $\delta G_N/G_N$ based on current CMB and BAO data do not depend significantly on the index of the monomial potential, but cosmological bounds on $\dot G_N/G_N$ 
do \cite{Ballardini:2016cvy}.

\begin{figure}
\includegraphics[width=.8\textwidth]{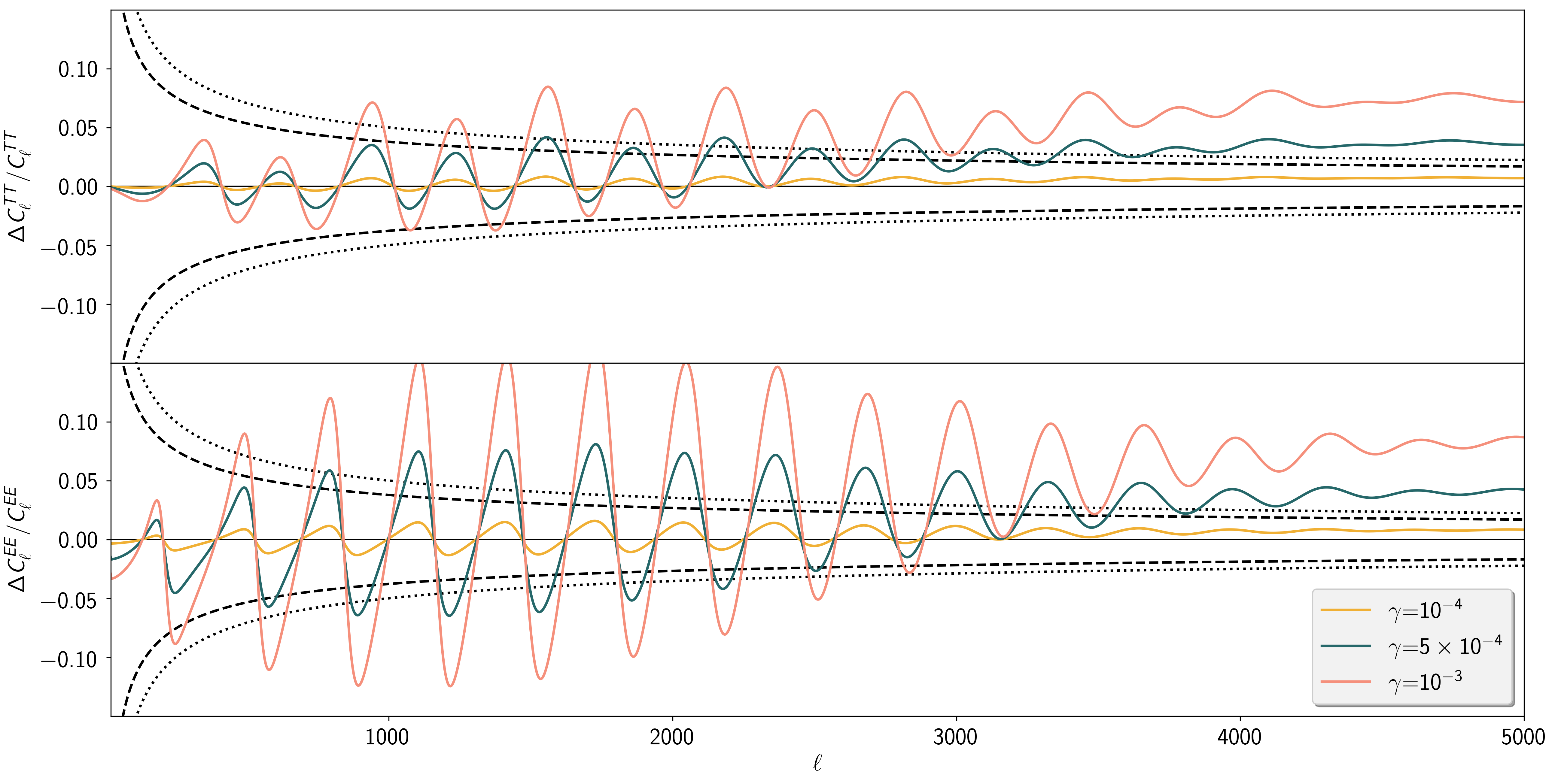}
\caption{Relative change in the CMB angular power spectra induced by different values 
of the coupling parameter from $\gamma=10^{-4}$ to $\gamma=10^{-3}$.
The black dashed (dotted) line refers to the noise spectrum of a cosmic-variance 
limited experiment with a sky fraction of $f_{\rm sky}=0.7\,(0.4)$.}
\label{fig:Cl-relative}
\end{figure}

There is still cosmological information for eJBD models to extract from the CMB pattern 
beyond $Planck$. 
In Fig.~\ref{fig:Cl-relative}, it shows the residuals of the lensed TT and EE CMB angular power 
spectrum as function of the multipole $\ell$ with respect to the sample variance for 
a sky fraction of $f_{\rm sky}=0.7$ and $f_{\rm sky}=0.4$.
Note the promise of E-mode polarization spectrum to constrain $\gamma$, 
we show the room of improvement on $\gamma$ expected from CMB temperature and polarization 
power spectra.

\section{Fisher approach for CMB anisotropy data}
\label{sec:three}
In this section, we start describing the formalism for our science forecasts. 
Under the Gaussian assumption for signal and noise, the Fisher matrix for CMB 
anisotropies in temperature and polarization  
\cite{Knox:1995dq,Jungman:1995bz,Seljak:1996ti,Zaldarriaga:1996xe,Kamionkowski:1996ks} is:
\be
\label{eqn:fisherCMB}
{\bf{\rm F}}_{\alpha\beta}^{\rm CMB} = f_{\rm sky}\sum_{\ell} \frac{(2 \ell+1) }{2}\sum_{\rm X,Y} \frac{\partial C_\ell^{\rm X}}{\partial\theta_\alpha}
\left({\rm Cov}_\ell^{-1}\right)_{\rm XY} \frac{\partial C_\ell^{\rm Y}}{\partial\theta_\beta} \,,
\ee
where $C_\ell^{\rm X}$ is the CMB angular power spectrum in the $\ell^\text{th}$ multipole 
for X,Y $\in ($TT$,\,$EE$,\,$TE$,\,\phi\phi,\,$T$\phi)$ 
\footnote{E$\phi$ has a negligible effect on the constraints, we do not consider its contribution.}, 
and $\theta_\alpha$ refers to the base of parameters considered in the analysis 
which are specified in Sec.~\ref{sec:five} togheter with their best-fit value.
The elements of the symmetric angular power spectrum covariance matrix $\text{Cov}_\ell$ 
at the $\ell^\text{th}$ multipole are:
\begin{align}
\left({\rm Cov}_\ell\right)_{\rm TTTT} 
&= \left(\bar{C}_\ell^{\rm TT}\right)^2-2\frac{\left(\bar{C}_\ell^{\rm TE}\bar{C}_\ell^{\rm T\phi}\right)^2}{\bar{C}_\ell^{\rm EE}\bar{C}_\ell^{\rm \phi\phi}} \,,\\
\left({\rm Cov}_\ell\right)_{\rm EEEE} 
&= \left(\bar{C}_\ell^{\rm EE}\right)^2 \,,\\
\left({\rm Cov}_\ell\right)_{\rm TETE} 
&= \frac{\left(\bar{C}_\ell^{\rm TE}\right)^2+\bar{C}_\ell^{\rm TT}\bar{C}_\ell^{\rm EE}}{2}
-\frac{\bar{C}_\ell^{\rm EE}\left(\bar{C}_\ell^{\rm T\phi}\right)^2}{2\bar{C}_\ell^{\rm \phi\phi}} \,,\\
\left({\rm Cov}_\ell\right)_{\rm \phi\phi\phi\phi} 
&= \left(\bar{C}_\ell^{\rm \phi\phi}\right)^2 \,,\\
\left({\rm Cov}_\ell\right)_{\rm T\phi T\phi} 
&= \frac{\left(\bar{C}_\ell^{\rm T\phi}\right)^2+\bar{C}_\ell^{\rm TT}\bar{C}_\ell^{\phi\phi}}{2}
-\frac{\bar{C}_\ell^{\phi\phi}\left(\bar{C}_\ell^{\rm TE}\right)^2}{2\bar{C}_\ell^{\rm EE}} \,,\\
\left({\rm Cov}_\ell\right)_{\rm TTEE} 
&= \left(\bar{C}_\ell^{\rm TE}\right)^2 \,,\\
\left({\rm Cov}_\ell\right)_{\rm TTTE} 
&= \bar{C}_\ell^{\rm TT}\bar{C}_\ell^{\rm TE}-\frac{\bar{C}_\ell^{\rm TE}\left(\bar{C}_\ell^{\rm T\phi}\right)^2}{\bar{C}_\ell^{\phi\phi}} \,,\\
\left({\rm Cov}_\ell\right)_{\rm TT\phi\phi} 
&= \left(\bar{C}_\ell^{\rm T\phi}\right)^2 \,,\\
\left({\rm Cov}_\ell\right)_{\rm TTT\phi} 
&= \bar{C}_\ell^{\rm TT}\bar{C}_\ell^{\rm T\phi}-\frac{\bar{C}_\ell^{\rm T\phi}\left(\bar{C}_\ell^{\rm TE}\right)^2}{\bar{C}_\ell^{\rm EE}} \,,\\
\left({\rm Cov}_\ell\right)_{\rm EETE} 
&= \bar{C}_\ell^{\rm EE}\bar{C}_\ell^{\rm TE} \,,\\
\left({\rm Cov}_\ell\right)_{\rm \phi\phi T\phi} 
&= \bar{C}_\ell^{\phi\phi}\bar{C}_\ell^{\rm T\phi} \,,
\end{align}
where $\bar{C}_\ell^{\rm X} = C^{\rm X}_\ell + N^{\rm X}_\ell$ is the sum of the signal and the noise, 
with $N^{\rm TE}_\ell,N^{\rm T\phi}_\ell=0$. 
For the temperature and polarization angular power spectra, here $N^{\rm X}_\ell=\sigma_{\rm X} b^{-2}_\ell$ 
is the isotropic noise convolved 
with the instrument beam, $b_\ell^2$ is the beam window function, assumed Gaussian, with 
$b_\ell = e^{-\ell (\ell + 1) \theta_{\rm FWHM}^2/16\ln 2}$; $\theta_{\rm FWHM}$ is the full 
width half maximum (FWHM) of the beam in radians; $w_{\rm TT}$ and $w_{\rm EE}$ are the inverse 
square of the detector noise level on a steradian patch for temperature and polarization, 
respectively. For multiple frequency channels, $\sigma^{-1}_{\rm X} b_\ell^2$ is replaced by sum 
of this quantity for each channels \cite{Knox:1995dq}:
\be
N^{\rm X}_\ell = \left[ \sum_{\rm channels} \frac{1}{N^{\rm X}_{\rm \ell,i}} \right]^{-1} \,.
\ee
We consider the minimum variance estimator for the noise of the 
lensing potential by combining the TT, EE, BB, TE, TB, EB CMB estimators 
calculated according to \cite{Hu:2001kj}.\\

In this paper, we consider four different cases as representative of current CMB 
measurements and future concepts.
We study the predictions for a $Planck$-like experiment consideridering the specifications 
of $f_{\rm sky} = 0.7$, and a multipole range from $\ell_{\rm min}=2$ up to $\ell_{\rm max} = 2500$ 
in Eq.~\eqref{eqn:fisherCMB}.
We use one cosmological frequency of 143 GHz assuming in flight performace corresponding to 
a sensitivity of $33\,\mu$K-arcmin in temperature and $70.2\,\mu$K-acmin 
in polarization, with a Gaussian beam width of 7.3 arcmin \cite{Adam:2015rua}, 
see CMB-1 in \cite{Ballardini:2016hpi}.

Since small-scale CMB anisotropy measurements will improve thanks to Stage-3 generation of 
ground-based CMB experiments, we consider AdvACT \cite{Calabrese:2014gwa,Allison:2015fac} with a 
noise level of $1.4\,\mu$K-arcmin in temperature and $8\,\mu$K-acmin in polarization, with a 
Gaussian beam width of 1.4 arcmin and $f_{\rm sky} = 0.4$, over a multipole range 
$30 \leq \ell \leq 3000$.

As concept for the next generation of CMB polarization experiments, we consider CORE and Stage-4 
(hereafter S4).
For CORE, we consider six frequency channels between 130 and 220 GHz with noise sensitivities 
of $1.5\,\mu$K-arcmin in temperature and $2\,\mu$K-acmin in polarization, with a Gaussian beam 
width of 5.5 arcmin \cite{Delabrouille:2017rct,DiValentino:2016foa,Finelli:2016cyd}. 
We consider $\ell_{\rm max}=3000$ for the CORE configuration with a sky coverage of $f_{\rm sky} = 0.7$.
 
The ground-based S4 proposal will be able to map modes up to $\ell\sim 5000$. 
Following \cite{Abazajian:2016yjj}, we consider for S4 a sensitivity 
$\sigma_{\rm T}=\sigma_{\rm P}/\sqrt{2}=1\,\mu$K-arcmin with a resolution of $\theta_{\rm FWHM}=3$ arcmin 
over $\sim40\%$ of the sky. Ground-based facilities are limited on large scales due to galactic 
foreground contamination and in addition a contamination is expected on the small scales in temperature. 
For these reasons, we assume for S4 $\ell_{\rm min}=30$ and a different cut at high-$\ell$ of 
$\ell_{\rm max}^{\rm T}=3000$ in temperature and $\ell_{\rm max}^{\rm P}=5000$ in polarization.
To complement at low multipoles, i.e. $2\leq \ell < 30$, we combine with $Planck$ AdvACT and 
with the Japan CMB polarization space mission proposal LiteBIRD \cite{Matsumura:2013aja,Errard:2015cxa} 
S4.
For the estimate of the noise of the lensing potential we use the multipole range 
$30 \leq \ell \leq 3000$.

\section{Fisher approach for LSS data}
\label{sec:four}
We now give the details for the Fisher forecasts with future LSS data.
We consider Euclid-like specifications as a representative case for future galaxy surveys. 
Euclid is a mission of the ESA Cosmic Vision program that it is expected to be launched in 2022. 
It will perform both a spectroscopic and a photometric survey: the first aims mainly at measuring the 
galaxy power spectrum of $\sim 30,000,000$ galaxies while the second at measuring 
the weak lensing signal by imaging $\sim 1.5$ billion galaxies.  

Both surveys will be able to constrain both the expansion and growth history of the universe and 
will cover a total area of $15,000$ square degrees.

\subsection{Spectroscopic galaxy power spectrum}

Following \cite{Seo:2003pu}, we write the linear observed galaxy power spectrum as: 
\be
P_{gal}(z:k_r,\mu_r) =
\frac{D_{Ar}^{2}(z)H(z)}{D_{A}^{2}(z)H_{r}(z)} \left[b(z)\sigma_8(z)+f(z,\,k)\sigma_8(z)\mu^{2}\right]^{2}  \frac{P_{r}(z,\,k)}{\sigma_8^2(z)}+ P_\mathrm{shot}(z) \,,
\label{eq:pk}
\ee
where the subscript $r$ refers to the reference (or fiducial) cosmological model.

Here $P_\mathrm{shot}(z)$ is a scale-independent offset due to imperfect removal of shot-noise, 
$\mu \equiv \vec{k}\cdot\hat{r}/k$ is the cosine of the angle of the wave mode with respect to 
the line of sight pointing into the direction $\hat{r}$, $P_{r}(z,\,k)$ is the fiducial matter 
power spectrum evaluated at different redshifts, $b(z)$ is the bias factor, $f(z,\,k)$ 
is the growth rate, $H(z)$ is the Hubble parameter  and $D_{A}(z)$ is the angular diameter distance. 
The wavenumber $k$ and $\mu$ have also to be written in terms of the fiducial cosmology 
(see for more details \cite{Seo:2003pu,Amendola:2004be,Sapone:2007jn}). 
The fiducial bias used in this paper is $b(z)=0.72 z+ 0.7$ according to \cite{Merson:2019vfr}.

The Fisher matrix for the galaxy power spectrum is given by \cite{Seo:2003pu}:
\be
F_{\alpha\beta} = \int_{k_\text{min}}^{k_\text{max}}\frac{k^2\dd k}{4\pi^2}\frac{\partial \ln P_{gal}(z;\,k,\mu)}{\partial \theta_\alpha}\frac{\partial \ln P_{gal}(z;\,k,\mu)}{\partial \theta_\beta}\times V_\text{eff}\,.
\ee
The observed galaxy power spectrum is given by Eq.~\eqref{eq:pk} and the derivatives are evaluated 
numerically at the fiducial cosmology; $k_\text{min} = 0.001\,h/\text{Mpc}$ and its value depends 
on the survey size whereas $k_\text{max}$ is such that root mean square amplitude of the density 
fluctuations at the scale $R_\text{max} = 2\pi/k_\text{max}\,\text{Mpc}/h$ is 
$\sigma^2(R_\text{max}) = 0.25$, however in order to not depend strongly on the non-linear information 
we consider two cases imposing an additional cut at $k_\text{max} = 0.1\,h/\text{Mpc}$ and at 
$k_\text{max} = 0.25\,h/\text{Mpc}$.
The effective volume of the survey in each bin is given by:
\be
V_\text{eff} = \left(\frac{\bar{n}\,P_{gal}(z;\,k, \mu)}{1+\bar{n}\,P_{gal}(z;\,k, \mu)}\right)^2V_\text{survey}\,, 
\ee
where $\bar{n}$ is the average comoving number density in each bin, the value of the $\bar{n}$ and fiducial specific Euclid-like specifications can be found in \cite{Pozzetti:2016cch,Merson:2017efv}.

To complete the GC information, we include low-redshift spectroscopic information 
from BOSS \cite{Dawson:2012va,Alam:2016hwk} on the redshift range $0.2 < z < 0.8$ over 10,000 
square degrees.

\subsection{Weak Lensing}
\label{subsec:wlps}

The weak lensing convergence power spectrum is given by 
\cite{Hu:1999ek,Hu:2002rm,Heavens:2003jx,Jain:2003tba,Amendola:2007rr}:
\begin{equation}
P_{ij}(\ell) = H_0^3
\int_0^{\infty}\frac{\dd z}{E(z)^2}\:W_{i}(z)W_{j}(z)\: P_\mathrm{NL}\left(k=\frac{H_0\,\ell}{r(z)},z\right) \,,
\label{eq:convergence-wl}
\end{equation}
where the subscript ${ij}$ refers to the redshift bins around $z_i$ and $z_j$, with $W_i(z)$ is 
the window function (see \cite{Majerotto:2015bra} for more details). 
The tomographic overall radial distribution function of galaxies for a Euclid-like 
photometric survey is \cite{Amendola:2012ys}:
\be
D(z) = z^2\exp\left[-\left(z/z_0\right)^{1.5}\right]\,.
\ee
with $z_0 = z_\text{mean}/1.412$ and mean redshift $z_\text{mean} = 0.9$, the number density 
is $d = 35 \text{ galaxy per arcmin}^2$. 
Moreover we consider a survey up to $z_\text{max}=3$ divided into 10 bins each containing the same 
number of galaxies. 
While tomography in general greatly reduces statistical errors the actual shape of the choice of 
the binning does not affect results in a serious way, although in principle there is room for 
optimisation \cite{Schaefer:2011dx}.

The Fisher matrix for weak lensing is defined as:
\be
F_{\alpha\beta} = f_\text{sky}\sum_\ell\frac{(2\ell+1)\Delta \ell}{2}\frac{\partial P_{i\,j}}{\partial\theta_\alpha}C_{j\,k}^{-1}\frac{\partial P_{k\,m}}{\partial\theta_\beta}C_{m\,i}^{-1} \,,
\ee
where $\Delta \ell$ is the step in multipoles, to which we chose 100 step in logarithm scale; whereas $\theta_\alpha$ are the cosmological parameters and:
\be
C_{j\,k} = P_{j\,k}+\delta_{j\,k}\langle\gamma_\text{int}^2\rangle\,n_j^{-1} \,,
\ee
where $\gamma_\text{int}$ is the rms intrinsic shear, which is assumed 
$\langle\gamma_\text{int}^2\rangle^{1/2} = 0.22$. 
The number of galaxies per steradians in each bin is defined as:
\be
n_j = 3600\,d\left(\frac{180}{\pi}\right)^2\hat{n}_j \,,
\ee
where $d$ is the number of galaxies per square arcminute and $\hat{n}_j$ is the faction of sources 
that belongs to the $j-$th bin.

\section{Statistical errors forecasts}
\label{sec:five}

In this section we estimate marginalised statistical errors for the cosmological 
parameters of our model, using the Fisher matrix calculation.
The probes are assumed to be independent, hence the total Fisher matrix is simply 
given by the sum of the single Fisher matrices: 
\be
F_{\alpha\beta} = F_{\alpha\beta}^\text{CMB}+F_{\alpha\beta}^\text{GC}+F_{\alpha\beta}^\text{WL} \,.
\ee
We perform the Fisher forecast analysis for the set of parameters 
$\omega_{\rm c},{\bf \theta} = \{\omega_{\rm b},h_0,n_{\rm s},\ln(10^{10}A_{\rm s}),\gamma\}$. 
For the CMB we consider also the reionization optical depth $\tau$ and then we marginalize over 
it before to combine the CMB Fisher matrix with the other two.
We assume as fiducial model a flat cosmology with best-fit parameters corresponding 
to $\omega_\textrm{c}\equiv\Omega_\textrm{c}h^2=0.1205$, 
$\omega_\textrm{b}\equiv\Omega_\textrm{b}h^2=0.02218$, $h_0\equiv H_0/100=0.6693$, $\tau=0.0596$, 
$n_\textrm{s}=0.9619$, and $\ln\left(10^{10}\ A_\textrm{s}\right)=3.056$ 
consistent with the recent results of $Planck$ \cite{Aghanim:2016yuo}.
As a fiducial value for the coupling to the Ricci curvature we choose $\gamma=10^{-5}$, 
the value is within the 95\% CL upper bound from current cosmological data \cite{Ballardini:2016cvy} 
and is contrained at 3$\sigma$ with the Solar System data \cite{Bertotti:2003rm}.
We have considered two fiducial potentials, a quartic potential and a constant one.

The CMB angular power spectra, the matter power spectra, together with the Hubble parameter $H(z)$, 
the angular diameter distance $D_A(z)$, and growth rate $f(z,k)$ have been computed with CLASSig, 
a modified version of the Einstein-Boltzmann code CLASS 
\footnote{\href{http://github.com/lesgourg/class_public}{http://github.com/lesgourg/class\_public}} 
\cite{Lesgourgues:2011re,Blas:2011rf} dedicated to eJBD theory \cite{Umilta:2015cta}.
This code has been successfully validated against other codes in \cite{Bellini:2017avd}. 
Non-linear scales have been included to the matter power spectrum assuming 
the halofit model from \cite{Takahashi:2012em}.
\begin{figure}
\includegraphics[width=.45\textwidth]{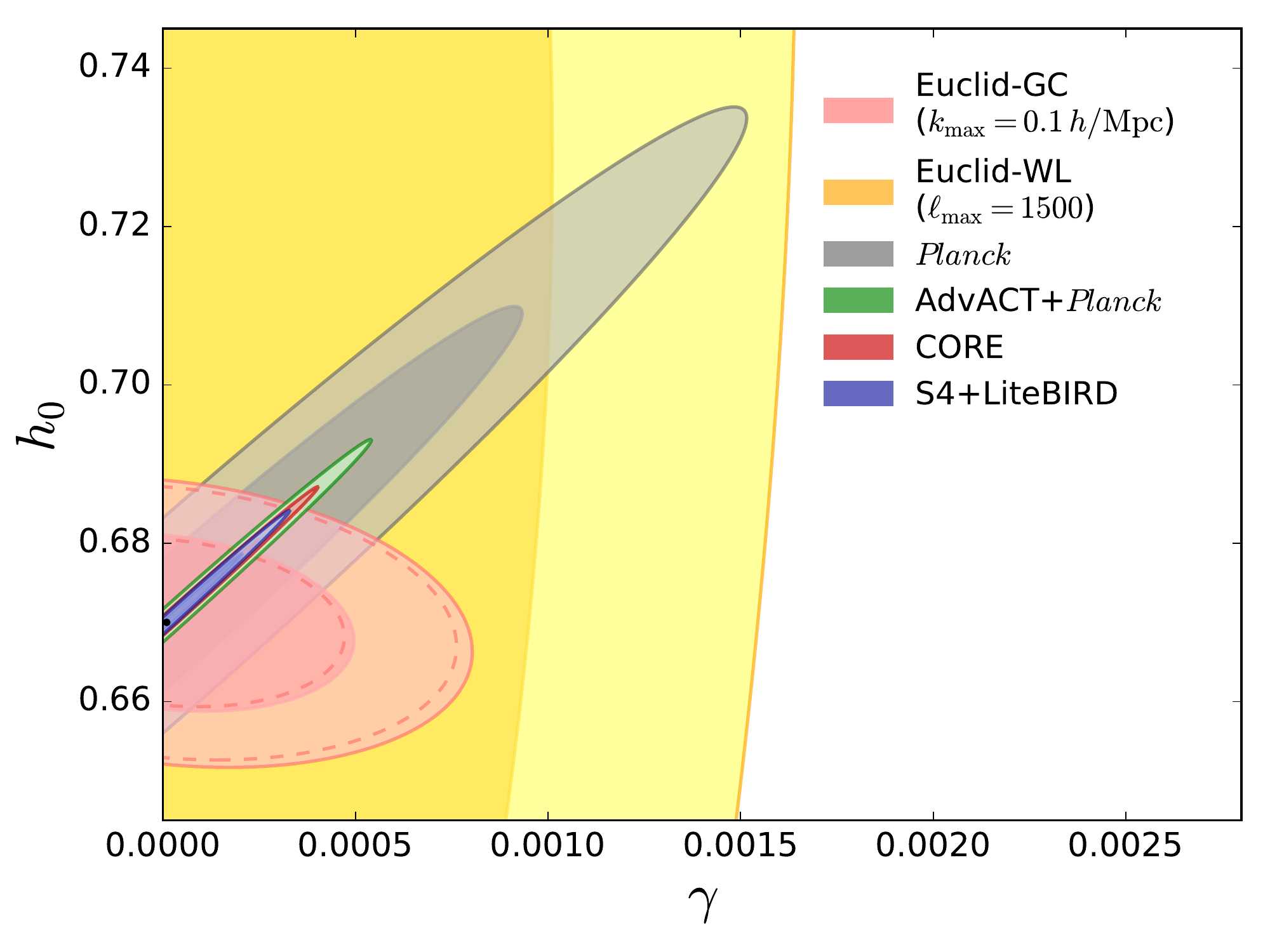}
\includegraphics[width=.45\textwidth]{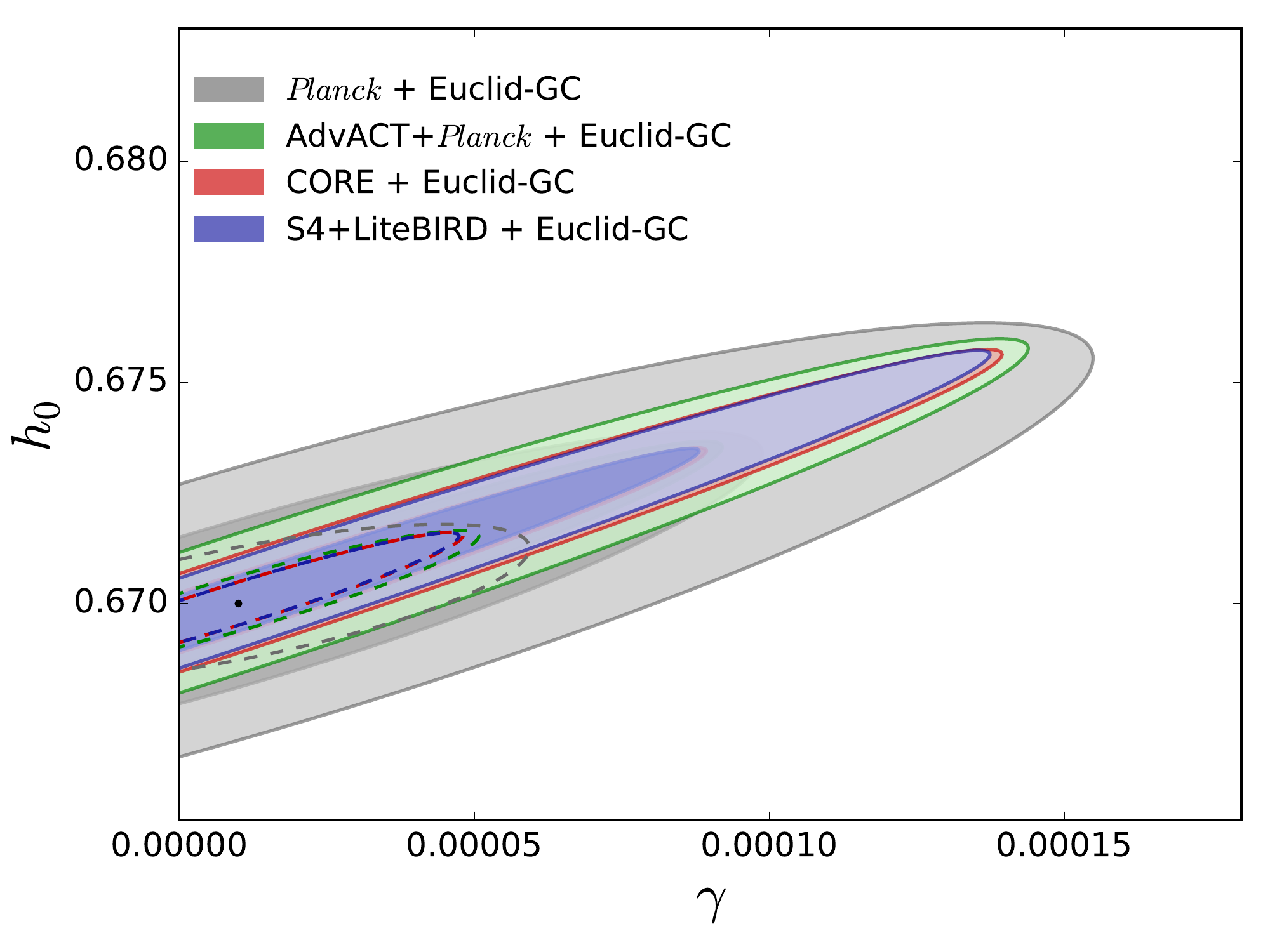}
\caption{Left: joint marginalized constraints (68\%-95\% CL) on $h_0$ and $\gamma$ from single 
probe alone. Dashed lines correspond to the 68\%-95\% CL using the GC information from Euclid-like 
in combination with BOSS. 
Right: joint marginalized constraints (68\%-95\% CL) on $h_0$ and $\gamma$ from the combination 
CMB+Euclid-GC for the four CMB surveys. Dashed lines correspond to the 68\% CL using the GC information 
up to $k_{\rm max}=0.25\,h/\text{Mpc}$.}
\label{fig:2D}
\end{figure}
Fig.~\ref{fig:2D} shows the constraints from the single observational probes. The different 
orientation of the 2-dimensional contours shows that the most efficient way to reduce the 
constraint error on $\gamma$ is to combine different cosmological probes.\\
With our Fisher approach, we find that the uncertainty from $Planck$ simulated data alone is 
$\sigma(\gamma)\simeq 0.00064$ at 68\% CL (consistent with our finding with $Planck$ 2015 real 
data \cite{Ballardini:2016cvy}) will improve by a factor three using  AdvACT+$Planck$, a factor 
four with CORE, and a factor five with the combination S4+LiteBIRD, i.e. 
$\sigma(\gamma)\simeq0.00022,\,0.00016,\,0.00013$ at 68\% CL respectively.

The combination of quasi-linear information with $k_{\rm max}=0.1\,h/\text{Mpc}$ from galaxy 
GC spectrum from Euclid-like and BOSS to the CMB leads to a significant improvement of the 
uncertainty on $\gamma$, approximately three-ten times with respect to the constraints obtained 
with the CMB alone.

In order to understand the improvement carried by mildly non-linear scales, we also include
the case of $k_{\rm max}=0.25\,h/\text{Mpc}$ which further improves the uncertainty on $\gamma$. 
In this case, we find five-twenty times better 
errors compared to CMB alone. We show in Fig.~\ref{fig:2D} the 2-dimensional 
marginal errors for the combination of CMB+Euclid-GC Fisher matricies for $Planck$, 
AdvACT+$Planck$, CORE, S4+LiteBIRD which correspond to the uncertainties of 
$\sigma(\gamma)\simeq0.000058\, (0.000032)$, 
$\sigma(\gamma)\simeq0.000054\, (0.000027)$, 
$\sigma(\gamma)\simeq0.000052\, (0.000025)$, 
$\sigma(\gamma)\simeq0.000051\, (0.000025)$ at 68\% CL for $k_{\rm max}=0.1\, (0.25)\,h/\text{Mpc}$; 
including also BOSS-GC information we obtain respectively
$0.000057\, (0.000031)$, 
$0.000052\, (0.000026)$, 
$0.000050\, (0.000024)$, 
$0.000049\, (0.000023)$ at 68\% CL for $k_{\rm max}=0.1\, (0.25)\,h/\text{Mpc}$.

Finally, we considered the combination of our three cosmological probes (CMB, GC, WL) to identify 
the tightest constraint on $\gamma$ by including non-linear scales through WL. 
The sensitivity on $\gamma$ combining the CMB with GC information up to 
$k_{\rm max}=0.1\,h/\text{Mpc}$ and WL assuming $\ell_{\rm max}=1500$ corresponds to 
$\sigma(\gamma)\simeq0.000045$, 
$\sigma(\gamma)\simeq0.000037$, 
$\sigma(\gamma)\simeq0.000029$, 
$\sigma(\gamma)\simeq0.000023$, 
at 68\% CL for $Planck$, AdvACT+$Planck$, CORE, and S4+LiteBIRD. These uncertainties 
improve of another 1.5 factor if we push the GC information up to $k_{\rm max}=0.25\,h/\text{Mpc}$.

\begin{figure}
\includegraphics[width=.45\textwidth]{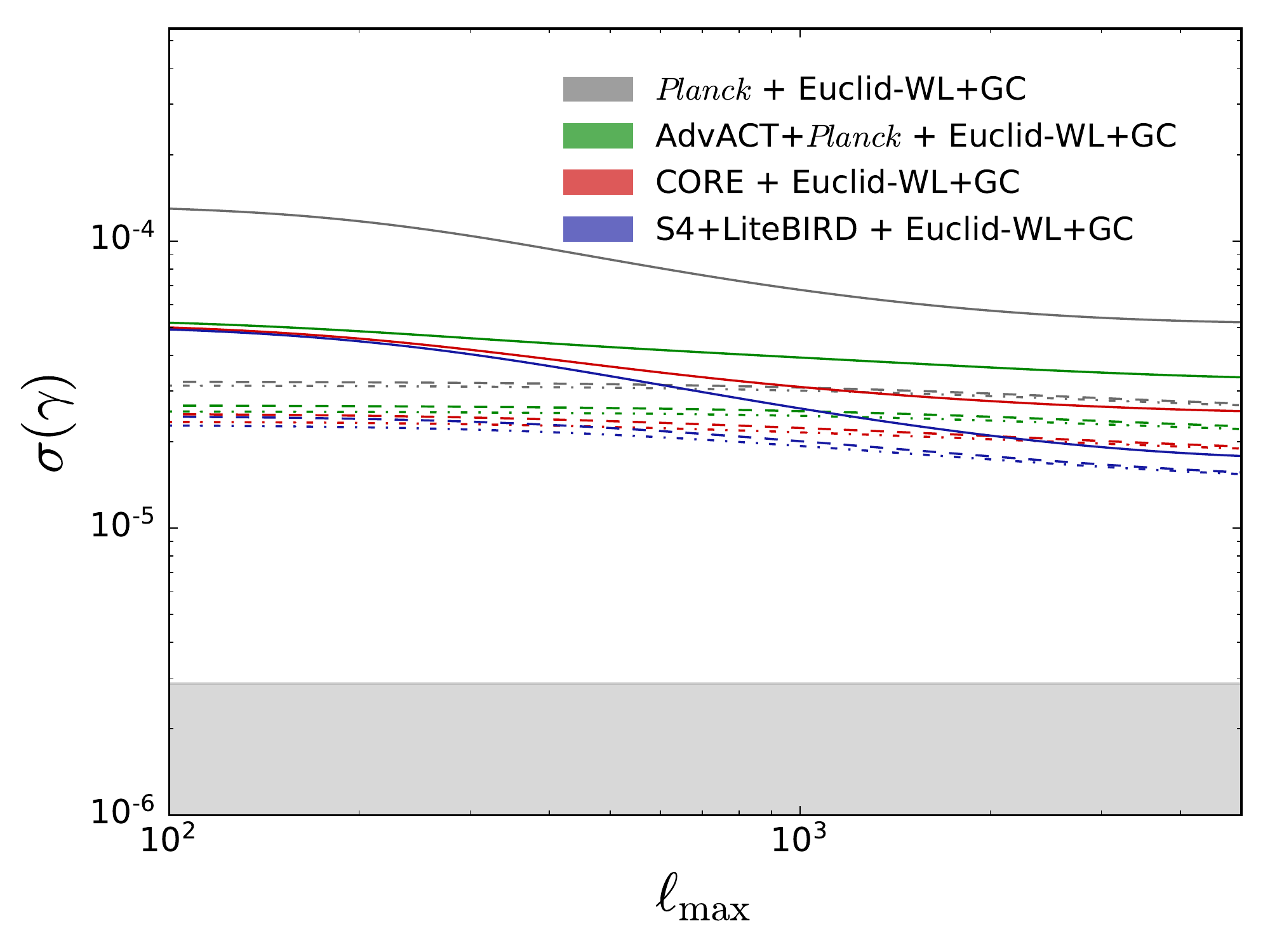}
\includegraphics[width=.45\textwidth]{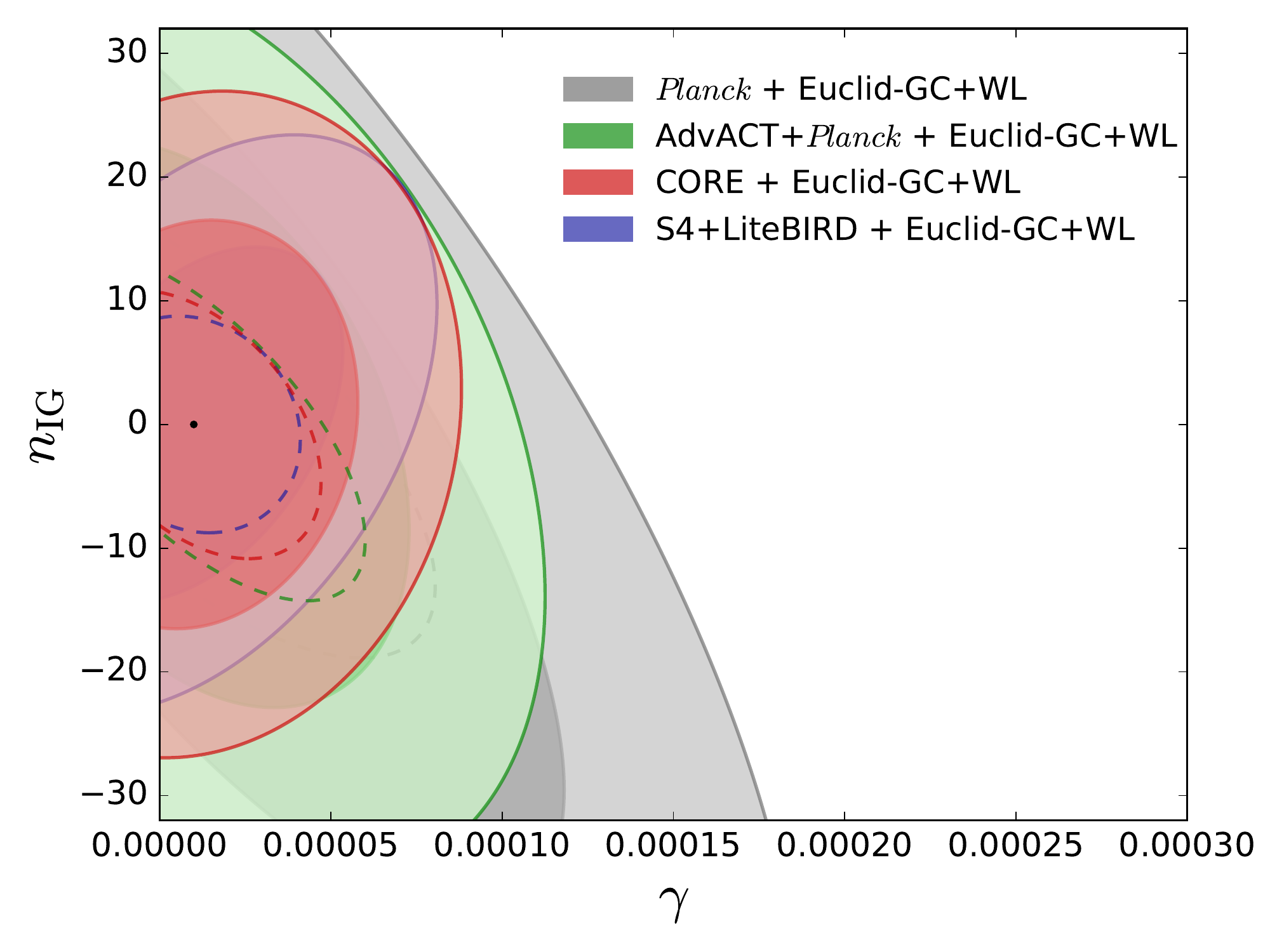}
\caption{Left: forecast marginalized constraint on $\gamma$ (68\% CL) as a function 
of the maximum multipole $\ell_{\rm max}$ included in the WL. Dashed lines 
correspond to the combination GC using the Euclid-GC information 
up to $k_{\rm max}=0.25\,h/\text{Mpc}$, and dot-dashed lines including also BOSS 
up to $k_{\rm max}=0.25\,h/\text{Mpc}$. The gray shaded 
region represent the 68\% CL contraint from Solar System data \cite{Bertotti:2003rm}. 
Right: joint marginalized constraints (68\%-95\% CL) on $n_{\rm IG}$ and $\gamma$ from the 
combination CMB+Euclid-GC for the four CMB surveys. Dashed lines correspond to the 68\% CL 
using the Euclid-GC information up to $k_{\rm max}=0.25\,h/\text{Mpc}$.}
\label{fig:wl-lmax}
\end{figure}

We show in Fig.~\ref{fig:wl-lmax} the impact of non-linear scales pushing the WL up to our optimistic 
case of $\ell_{\rm max}=5000$. We find a small improvement in terms of error on $\gamma$ 
pushing $\ell_{\rm max}$ for the WL from 1500 to 5000.

The most conservative forecast of Euclid-like plus BOSS to quasi-linear scales in combination 
with $Planck$ improve between a factor three and approximately a factor twenty the current 
uncertainties based on $Planck$ 2015 and BAO data.

In Tab.~\ref{tab:errors}, we show the marginalized uncertainties on all the 
cosmological parameters from the combination of the CMB surveys with Euclid-like and 
BOSS using the conservative range of scales.

\begin{table}[!h]
\centering
\caption{Marginalized uncertainties (68\% CL) on the cosmological parameters 
for $\gamma=10^{-5}$ and $n_{\rm IG}=4$. 
We consider the combination of three CMB surveys with the information up 
to quasi-linear scales from the GC ($k_{\rm max}=0.1\ h/$Mpc) using Euclid-like 
plus BOSS and WL ($\ell_{\rm max}=1500$) from Euclid-like.
Numbers in round brakets refer to the uncertainties for our optimistic case 
with GC up to $k_{\rm max}=0.25\ h/$Mpc and WL up to $\ell_{\rm max}=5000$.}
\label{tab:errors}
\vspace*{0.2cm}
\begin{tabular}{|c|ccc|}
\hline
\rule[-1mm]{0mm}{.4cm}
 & $Planck$ & AdvACT+$Planck$ & S4+LiteBIRD \\
 & + BOSS-GC & + BOSS-GC & + BOSS-GC \\
 & + Euclid-GC+WL & + Euclid-GC+WL & + Euclid-GC+WL \\
\hline
\rule[-1mm]{0mm}{.4cm}
$10^4~\sigma(\omega_c)$ & 1.3 (1.1) & 1.0 (0.78) & 0.83 (0.67) \\
\rule[-1mm]{0mm}{.4cm}
$10^5~\sigma(\omega_b)$ & 9.8 (8.8) & 4.5 (3.8) & 3.2 (2.8) \\
\rule[-1mm]{0mm}{.4cm}
$10^3~\sigma(h)$ & 1.7 (0.99) & 1.4 (0.84) & 1.0 (0.64) \\
\rule[-1mm]{0mm}{.4cm}
$10^3~\sigma(n_s)$ & 1.8 (1.2) & 1.5 (0.96) & 1.2 (0.88) \\
\rule[-1mm]{0mm}{.4cm}
$10^3~\sigma\left(\ln\left(10^{10}A_s\right)\right)$ & 1.3 (0.75) & 1.1 (0.71) & 0.98 (0.56) \\
\rule[-1mm]{0mm}{.4cm}
$10^5~\sigma(\gamma)$ & 4.5 (2.7) & 3.7 (2.2) & 2.3 (1.5) \\
\hline
\end{tabular}
\end{table}

We study the impact of a larger fiducial value for $\gamma=10^{-4}$ 
(still compatible with current $Planck$ 2015 + BAO constraints) on the forecasted uncertainties. 
We find that the effect on CMB and GC is around $\sim 10\%$ on the uncertainties; this implies 
that we will be able to detect at 2-5~$\sigma$ CL a value of $\gamma=10^{-4}$ with the combination 
of the CMB with GC information from a Euclid-like experiment. 
Regarding the WL, the uncertainties halve leading to a clearer detection of such $\gamma$ at more 
then 5~$\sigma$ when WL from Euclid-like is added.

We repeat our series of forecasts with a different potential for the scalar field with an index 
equal to zero, i.e. $n_{\rm IG}=0$, namely a cosmological constant.
For this fiducial cosmology we find only a small degradation of the uncertainties on $\gamma$, 
pointing to a weak correlation between the two parameters.

Finally, we test also the possibility to constraint the index of the scalar potential 
around $n_{\rm IG}=0$ with future cosmological data, see Fig.~\ref{fig:wl-lmax}. 
The tightest uncertainty that we obtain combining all the three cosmological probes and 
including non-linear scales in the GC up to $k_{\rm max}=0.25\,h/\text{Mpc}$ and in the 
WL up to $\ell_{\rm max}=5000$ is $\sigma\left(n_{\rm IG}\right) \simeq 5$ at 68\% CL.

In order to compare our finding on cosmological scales with the constraints obtained within 
the Solar System we quote the constraint on the post-Newtonian parameters defined for this 
class of eJBD theories as:
\be
\gamma_{\rm PN} = \frac{1+4\gamma}{1+8\gamma} \,.
\ee
Our derived forecasted uncertainties span between 
$\sigma\left(|\gamma_{\rm PN}-1|\right) \simeq 5.3\times10^{-4}-2.5\times10^{-3}$ at 68\% CL 
for a CMB experiment with a $Planck$ sensitivity through a future CMB experiment 
able to perform a cosmic-variance measurement of the E-mode polarization at small scales.
Combining CMB information with GC and WL, we find 
$\sigma\left(|\gamma_{\rm PN}-1|\right) \simeq 9.2\times10^{-5}$ and including non-linear 
scales a minimum error on the deviation from GR in the weak field limit corresponding 
to $\sigma\left(|\gamma_{\rm PN}-1|\right) \simeq 6.2\times10^{-5}$ at 68\% CL.

\section{Conclusion}
\label{sec:conclusion}
eJBD theories represent the simplest scalar-tensor theory of gravity where 
Newton's constant is allow to vary becoming a dynamical field, as a function 
of space and time.

This class of theories have been already severely constrained from Solar System experiments 
leading to $\gamma_\mathrm{PN} - 1 = (2.1 \pm 2.3) \times 10^{-5}$ at 68\% CL \cite{Bertotti:2003rm}. 
These Solar System tests constrain the weak-field behaviour of 
gravity, and the strong-field behaviour that this family of theories can still 
exhibit is contrained by the binary pulsar \cite{Zhu:2015mdo,Archibald:2018oxs}.

However, it is conceivable that gravity differed considerably from GR in the early Universe. 
Even if GR seems to work well today on Solar System scales, 
in several scalar-tensor theories there is generally an attractor mechanism that
drives to an effective cosmological constant at late time.
BBN \cite{Copi:2003xd,Bambi:2005fi} provides a test of gravity at 
early times based on the impact of the effective gravitational constant on the expansion rate 
and on the cosmological abundances of the light elements produced during BBN.

Cosmological observations, such as CMB anisotropies and LSS matter distribution,  
probe different epochs and scales of the Universe.
The redshift of matter-radiation equality is modified in eJBD theories by the motion 
of the scalar field driven by pressureless matter and this results in a shift 
of the CMB acoustic peaks \cite{Liddle:1998ij,Chen:1999qh}.

$Planck$ data have been already used to constrain this eJBD models 
\cite{Avilez:2013dxa,Li:2013nwa,Umilta:2015cta,Ooba:2016slp} 
(see \cite{Nagata:2003qn,Acquaviva:2004ti,Wu:2009zb} for analysis with pre-$Planck$ data). 
Latest $Planck$ 2015 publicly data release constrain $1 - \gamma_\mathrm{PN} < 0.007$ at 95\% CL,
and the combination of CMB and LSS data through the addition of BAO information 
has shown a promising way to further constrain these class of models in light of upcoming 
LSS experiments, leading to $1 - \gamma_\mathrm{PN} < 0.003$ at 95\% CL \cite{Ballardini:2016cvy}.

In this paper we investigated how well some future CMB experiments and LSS surveys will be able 
to improve current cosmological constraints on this simple scalar-tensor theories.
We consider eJBD theory of gravity where a potential term is included in order to embed in the 
original JBD theory the current acceleration phase of the Universe.
Our results have been enlightening and we can summarise them as follows:
\begin{itemize}
\item Future CMB experiment, such as AdvACT, CORE and Stage-4 CMB, will improve current 
constraints from $Planck$ 2015 alone by a factor 3-5, thanks to a better measure of the small 
scale CMB anisotropies. We find that in the best case $\sigma\left(1 - \gamma_\mathrm{PN}\right) \simeq 0.0005$ at 68\% CL 
for propose space and ground-base CMB experiment CORE and S4.
\item We forecast the combination of CMB and spectroscopic surveys using the 3-dimensional 
observed galaxy power spectrum. We consider a Euclid-like spectrocopic survey and  
to complete the redshift coverage of a Euclid-like selection function $0.9<z<1.8$ 
\cite{Pozzetti:2016cch,Merson:2017efv}
we include optical spectroscopic observations from BOSS in the range $0.2<z<0.8$ 
\cite{Dawson:2012va,Alam:2016hwk}.

The combination of quasi-linear information up to $k_{\rm max}=0.1\,h/\text{Mpc}$ for the 
Euclid-like and BOSS GC to the CMB leads roughly a reduction around three-ten times with respect 
to the uncertainties obtained 
with the CMB alone, with a best case bound of $\sigma\left(1 - \gamma_\mathrm{PN}\right) \simeq 0.0002$ at 68\% CL.
\item We find that the inclusion of mildly non-linear scales in the galaxy power spectrum 
is crucial to drive the contraints from cosmological observation at the same order of current 
Solar System constraints. 
\item WL surves will improve the sensitivity on $\gamma$ approximately by a factor 2.
\item The best bound that we obtain combining all the three cosmological probes and including 
non-linear scales in the GC up to $k_{\rm max} = 0.25\,h/\text{Mpc}$ and in the WL up to 
$\ell_{\rm max} = 5000$ is $\sigma\left(1 - \gamma_\mathrm{PN}\right) \simeq 0.000062$ at 68\% CL.

This forecast is only approximately a factor three worst than the current Solar System constraint.
\end{itemize} 

Although consistent with \cite{Acquaviva:2007mm,Alonso:2016suf} our estimate of $\gamma_{\rm PN}$ is 
based on different assuptions.
It is difficult to compare our results with the pioneering work \cite{Acquaviva:2007mm}: 
theoretical predictions, forecast methodology and experimental specifications in \cite{Acquaviva:2007mm} 
are different from our analysis. Overall, we can say that our forecasted uncertainty on $\gamma_{\rm PN}$ is 
more optimistic than those quoted in \cite{Acquaviva:2007mm} because we combine expected constraints from different probes. 
\cite{Alonso:2016suf} use LSST photometric specifications for galaxy clustering and weak lensing, 
and SKA1-MID intensity mapping, whereas we use Euclid-like spectroscopic survey for galaxy clustering 
and photometric specifications for weak lensing; we do not consider any screening.

We close with three final remarks. Our forecasted sensitivity on $\gamma_{\rm PN}$ is smaller 
than the one obtained from models with a non-universal coupling between dark 
matter and dark energy and motivated by eJBD theories \cite{Amendola:2011ie}. As second remark, 
our works shows the importance of developing non-linear approximation schemes for eJBD theories 
\cite{Perrotta:2003rh,Li:2010zw,Taruya:2014faa,Winther:2015wla} to reach the accuracy required 
by future cosmological observations. As a third and conclusive point, it would be interesting to 
further add complementary probes at low redshift: indeed, we have been quite inclusive with the 
forecasts from next CMB polarisation experiments whereas other measurement at lower redshift 
complementary to Euclid and BOSS and might be crucial to strengthen our predictions.

\section*{Acknowledgements}
MB was supported by the South African Radio Astronomy Observatory, which is a 
facility of the National Research Foundation, an agency of the Department of Science 
and Technology and he was also supported by the Claude Leon Foundation.
DS acknowledges financial support from the Fondecyt project number 11140496.
DS would like to thank INFN for supporting a visit in Bologna during which this work was carried on.
This work has made use of the Horizon Cluster hosted by Institut d'Astrophysique de Paris. 
We thank Stephane Rouberol for smoothly running this cluster. UC was partially supported 
within the Labex ILP (reference ANR-10-LABX-63) part of the Idex SUPER, and received 
financial state aid managed by the Agence Nationale de la Recherche, as part of the 
programme Investissements d'avenir under the reference ANR-11-IDEX-0004-02.
MB, DP and FF acknowledge financial support by ASI n.I/023/12/0 "Attivit\`a relative alla 
fase B2/C per la missione Euclid", ASI Grant 2016-24-H.0 and partial financial 
support by the ASI/INAF Agreement I/072/09/0 for the Planck LFI Activity of Phase E2.




\end{document}